\documentclass[twocolumn,longbibliography,aps,pra,preprintnumbers,10pt]{revtex4-1}

\usepackage{graphicx}
\usepackage{bm}
\usepackage{color}
\usepackage{amsmath}
\usepackage{amssymb}
\usepackage{enumerate}

\usepackage[urlcolor=blue,colorlinks=true,citecolor=blue,linkcolor=blue,pdfstartview={FitH},bookmarks=false]{hyperref}

\graphicspath{{fig_new/}{./fig_new/}{.}}

\begin{document}

\title{Structural, electronic, and dynamical properties \\ of the tetragonal and collapsed tetragonal phases of KFe$_2$As$_2$}

\author{Andrzej Ptok}
\email[e-mail: ]{aptok@mmj.pl}
\affiliation{Institute of Nuclear Physics, Polish Academy of Sciences, \\
ul. W. E. Radzikowskiego 152, PL-31342 Krak\'{o}w, Poland}

\author{Ma\l{}gorzata Sternik}
\email[e-mail: ]{malgorzata.sternik@ifj.edu.pl}
\affiliation{Institute of Nuclear Physics, Polish Academy of Sciences, \\
ul. W. E. Radzikowskiego 152, PL-31342 Krak\'{o}w, Poland}

\author{Konrad Jerzy Kapcia}
\email[e-mail: ]{konrad.kapcia@ifj.edu.pl}
\affiliation{Institute of Nuclear Physics, Polish Academy of Sciences, \\
ul. W. E. Radzikowskiego 152, PL-31342 Krak\'{o}w, Poland}

\author{Przemys\l{}aw Piekarz}
\email[e-mail: ]{piekarz@wolf.ifj.edu.pl}
\affiliation{Institute of Nuclear Physics, Polish Academy of Sciences, \\
ul. W. E. Radzikowskiego 152, PL-31342 Krak\'{o}w, Poland}

\begin{abstract}
Compounds with a tetragonal ThCr$_{2}$Si$_{2}$-type structure are characterized by the possibility of the isostructural phase transition from the tetragonal phase to the collapsed tetragonal phase induced by the external pressure.
An example of a compound with such a phase transition is KFe$_{2}$As$_{2}$, which belongs to the family of the iron-based superconductors.
In this paper, we investigate the effects of the phase transition on the structural, electronic, and dynamical properties of this compound.
Performing the {\it ab initio} calculations, we reproduce the dependence of the lattice constants on pressure and analyze the changes of the inter-atomic distances
in the tetragonal and collapsed tetragonal phases. 
Using the tight binding model with maximally localized Wannier orbitals, we calculate and discuss the influence of pressure on the electronic band structure as well as on the shape of the Fermi surface.
We found a precursor of the phase transition in the form of enhancement of overlapping between two Wannier orbitals of As atoms.
In order to better understand the superconducting properties of KFe$_{2}$As$_{2}$, we study the orbital-projected Cooper pairs susceptibility  as a function of pressure. 
We found a decrease of susceptibility with the increasing pressure in a good qualitative agreement with experimental observation.
The structural transition also influences the phonon spectrum of KFe$_{2}$As$_{2}$, which exhibits pronounced changes induced by pressure.
Some phonon modes related with the vibrations of Fe and As atoms show an anomalous, nonmonotonic dependence on pressure close to the phase transition.
\end{abstract}

\maketitle

\section{Introduction}
\label{sec.intro}

A discovery of the high temperature iron-based superconductors (IBSC) in 2008~\cite{kamihara.watanabe.08} opened a period of intensive studies of these compounds~\cite{stewart.11,dagotto.13,dai.15,hosono.akiyasu.17}.
One of the most investigated family is {\it A}Fe$_2$As$_2$ (or shortly {\it A}122), where {\it A} is an alkali metal or rare-earth metal.
In this family, a heavily hole-doped K122, which crystallizes in the tetragonal ThCr$_{2}$Si$_{2}$-type structure ({\it I4/mmm}, space group 139), is particularly interesting. 
It is the end-member of the Ba$_{1-x}$K$_{x}$122 series, where the maximum $T_{c}=38$~K is achieved at optimal doping $x \sim 0.4$~\cite{rotter.tegel.08}.
K122 with $T_{c} \sim$ 3.5~K~\cite{sato.nakayama.09,fukazawa.yamada.09}, in contrary to the Ba122 compound, does not show any magnetic order~\cite{avci.chmaissem.12}.

In a general case the IBSC exhibit a structural phase transition from the tetragonal phase to the orthorhombic one due to changes in temperature or chemical pressure~\cite{wang.lee.11}.
K122 crystallizes in the tetragonal phase and under external pressure exhibits a phase transition from the tetragonal (T) to the collapsed tetragonal (cT) phase, with the same symmetry but with strongly and not uniformly modified lattice parameters.
The T--cT phase transition has been reported not only for the IBSC (like e.g. K122~\cite{nakajima.wang.15,ying.tang.15}, Ca122~\cite{kreyssig.green.08}, Ba122~\cite{uhoya.stemshorn.10,mittal.mishra.11} and Eu122~\cite{uhoya.tsoi.10}), but also for other compounds with the ThCr$_{2}$Si$_{2}$-type structure (e.g. SrRh$_{2}$P$_{2}$~\cite{johrendt.felser.97}, CaFe$_{2}$P$_{2}$~\cite{coldea.andrew.09}, EuFe$_{2}$P$_{2}$~\cite{huhnt.schlabitz.98}, SrCo$_{2}$As$_{2}$~\cite{jayasekara.kaluarachchi.15} or BaCr$_{2}$As$_{2}$~\cite{naumov.filsinger.17}).

External pressure changes the superconducting properties of the K122 tetragonal phase.
$T_{c}$ shows a non-monotonic behavior on pressure~\cite{taufour.foroozani.14,ying.tang.15,nakajima.wang.15,wang.zhou.16,wang.matsubayashi.16,tafti.juneaufecteau.13}, with a minimum at $p_{c} \sim$ 1.8 GPa~\cite{tafti.juneaufecteau.13}.
Around this point, $T_{c}$ shows a universal V--shape dependence in the weak pressure regime~\cite{tafti.ouellet.15}, what has been reported not only in the case of K122~\cite{tafti.juneaufecteau.13}, but also for Rb122~\cite{tafti.ouellet.15} and Cs122~\cite{tafti.clancy.14}.
Early experimental studies at higher pressures report two different superconducting domes around the structural phase transition with the enhanced critical temperature in the collapsed tetragonal phase ~\cite{nakajima.wang.15}.
The band structure calculations predicted the occurrence of the Lifshitz transition and the change of gap symmetry from {\it d-wave} to {\it $s_{\pm}$-wave} when entering the cT phase~\cite{guterding.backes.15}.
However, more recent studies indicated a gradual suppression of superconductivity under pressure with $T_{c}$ going to zero at $\sim11$ GPa and an absence of superconductivity in the cT phase~\cite{wang.matsubayashi.16}.

The isostructural phase transition also influences the dynamical properties of 122 crystals. 
Significant changes in the atomic vibrations, mainly along the $z$ direction, have been observed in high-pressure measurements using the inelastic x-ray and neutron scattering for Ca122~\cite{mittal.heid.09} and the nuclear resonant inelastic scattering for Sr122~\cite{wang.lu.16}.
One should expect similar effects in K122, however, the phonon properties of this compound have not been studied yet.

In this paper, in order to obtain a deeper insight in the structural phase transition from the T phase to the cT phase, structural, electronic, and dynamical properties of K122 under hydrostatic pressure have been studied using the density functional theory (DFT).
We analyze in details changes in the electron band structure and discuss the Lifshitz transition induced by pressure. 
Using the tight binding model (TBM) constructed within  maximally localized Wannier (MLW) functions, we study the effect of pressure on the pairing susceptibility
and discuss the mechanism that leads to suppression of superconductivity in K122.  
Our studies can be generalized to other members of the $122$ family and help to explain the partial cT phase in a new IBSC 1144 family~\cite{iyo.kawashima.16,song.nguyen.18,borisov.canfield.18} (e.g. CaKFe$_{4}$As$_{4}$~\cite{mou.kong.16,kaluarachchi.taufour.17}, RbEuFe$_{4}$As$_{4}$ and CsEuFe$_{4}$As$_{4}$~\cite{jackson.vangennep.18}).
The present paper is organized as follows.
In Section~\ref{sec.numres}, we show and describe the influence of the external hydrostatic pressure on physical properties of the system, especially: the main structural properties (Sec.~\ref{sec.collapse}), electronic properties (Sec.~\ref{sec.electrons}), superconducting properties~\ref{sec.supercond}, and dynamical properties (Sec.~\ref{sec.phonons}).
Finally, we summarize the results in Section~\ref{sec.summary}.

\begin{figure*}[!t]
\includegraphics[width=0.8\linewidth]{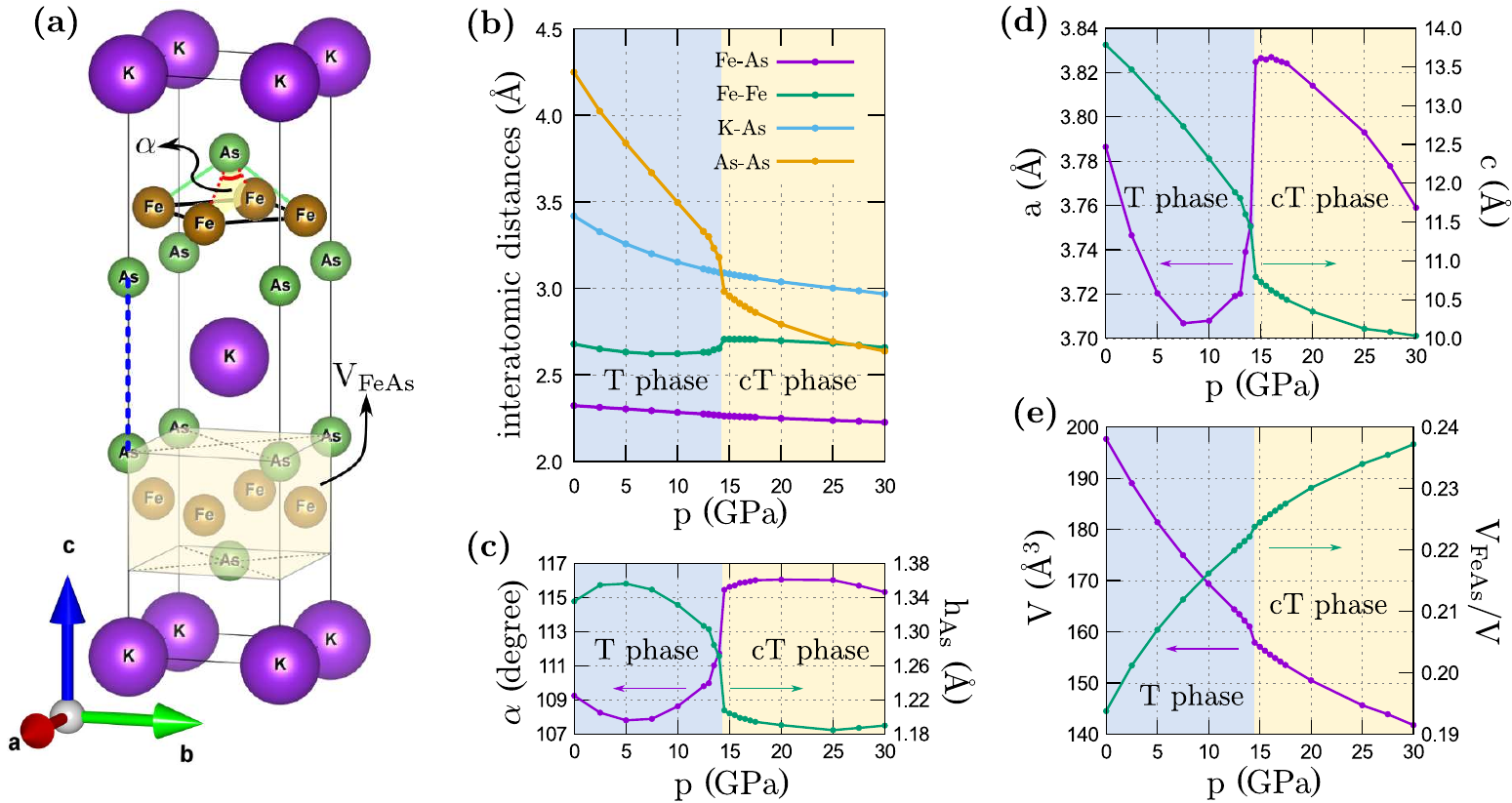}
\caption{
\label{fig.struc}
(a) The conventional cell of the tetragonal ({\it I4/mmm}) structure of KFe$_2$As$_2$. 
The image was rendered using {\sc VESTA} software~\cite{vesta}.
Pressure dependences of the following structural parameters: 
(b) Fe--As, Fe--Fe, K--As, and As--As (as labeled) interatomic distances, (c) angle $\alpha$ between two Fe--As bonds [shown in panel (a)] and distance h$_{\text{As}}$ of As atom from Fe--plane,
(d) lattice constants $a$ and $c$,
and 
(e) volume $V$ of the K122 crystal cell (in a conventional standard) and relative volume $V_{\text{FeAs}}/V$ of Fe--As layer [shown in panel (a)].
}
\end{figure*}

\section{Numerical results and discussion}
\label{sec.numres}

Main DFT calculations were performed using the projector augmented-wave method~\cite{PAW}, within the generalized gradient approximation~\cite{PBE} to the exchange-correlation functional, implemented in the {\sc VASP} program~\cite{VASP1,VASP2}.
The method of Methfessel-Paxton broadening technique~\cite{methfessel.paxton.89} with $0.2$~eV width of smearing was adopted to describe the partial occupancies for each wave function.
A ${\bm k}$-mesh of $16 \times 16 \times 8$ points in the Monkhorst-Pack scheme~\cite{monkhorst.pack.76} was used for  integration in the reciprocal space and the energy cut-off for plane wave expansion of $450$~eV was applied.
The crystal structure was optimized using the conjugate gradient technique with the energy convergence criteria set at $10^{-7}$~eV and $10^{-5}$~eV for electronic and ionic iterations, respectively, in the conventional cell.
During calculations, the lattice constants $a$ and $c$ as well as the position of As atoms ($z_{As}$) were optimized.
The optimization procedure was repeated for various hydrostatic pressures imposed on the K122 structure. 
The numerical results and characterization of the optimized parameters will be shown in Sec.~\ref{sec.collapse}.


Next, for the optimized structures, we present the electronic properties of the system in the normal phase, i.e., not superconducting one (Sec.~\ref{sec.electrons}).
Additionally, we predict properties of the K122 in the superconducting state by using the Cooper pairs susceptibility, described in details in Ref.~\cite{ptok.kapcia.17}.
To do this, we express the Bloch states in the form of the TBM using the MLW orbitals~\cite{marzari.mostofi.12}.
These calculations were performed by the {\sc Wannier90} program~\cite{mostofi.yates.08,mostofi.yates.14}.
We chose {\it d} orbitals centered at Fe atoms and {\it p} orbitals centered at As atoms building the $16$-orbitals TBM.
The calculations were performed in a primitive unit cell.
The TBM was also used to find the shape of the Fermi surfaces (FS) on a $50 \times 50 \times 50$ ${\bm k}$-grid and orbital projections.


In the last part, the phonon calculations are presented (Sec.~\ref{sec.phonons}).
The dynamical properties of K122 were calculated using the direct force constant approach (the Parlinski--Li--Kawazoe method)~\cite{phonon1} implemented in the {\sc Phonon} program~\cite{phonon2}.
In these calculations, we used the supercell containing $2 \times 2 \times 2$  conventional cells, with simultaneous reduction of the ${\bm k}$-mesh to $4 \times 4 \times 2$.
The force constants were obtained from first-principle calculations of the Hellmann-Feynman forces by {\sc VASP} and they were used to build a dynamical matrix of the crystal. 
Phonon frequencies were obtained by the diagonalization of the dynamical matrix.

\subsection{Isostructural phase transition}
\label{sec.collapse}

As it was mentioned, K122 crystallizes in a tetragonal ThCr$_{2}$Si$_{2}$-type structure ({\it I4/mmm}, space group: $139$), which consists of Fe$_2$As$_2$ layers separated by potassium layers. 
With increasing hydrostatic pressure, the phase transition from T to cT structure occurs around 16~GPa.
At this transition the crystal structure does not change its symmetry.
In both phases, the primitive unit cell contains single KFe$_2$As$_2$ chemical formula, but it is more convenient to use the crystallographic cell, which is shown in Fig.~\ref{fig.struc}(a).
In this cell three non-equivalent atoms K, Fe, and As are placed at the crystallographic sites: $2a(0,0,0)$, $4d(0.5,0,025)$ and $4e(0,0,{ z}_{As})$, respectively, defining all remaining atomic positions.
In the absence of pressure, the structural parameters of the K122 tetragonal phase were specified as $a = b = 3.842$~\AA, $c = 13.861$~\AA, and $z_{As}=0.3525$~\cite{sandor.uwe.81}.
Within our numerical calculations we found $a = b = 3.782$~\AA, $c = 13.876$~\AA, and $z_{As}=0.3469$. 

The pressure dependence of the optimized structural parameters (distances between atoms, an angle between two Fe--As bonds, lattice constants as well as volumes of the crystallographic cell and the FeSe layer) are presented in Fig.~\ref{fig.struc}(b)--(e).
The differences beetween the theoretical lattice constants and the experimental data~\cite{nakajima.wang.15} in the entire range of pressure are smaller than 1.5\%. This is a typical error in DFT calculations and it does not depend much on pressure.
Therefore, our structural studies under pressure are reliable and sufficiently accurate to reproduce the relationship between the lattice parameters and 
pressure observed experimentally by the diffraction studies~\cite{nakajima.wang.15}.
This can clearly be demonstrated by comparing Fig.~\ref{fig.struc}(d) with Fig.3(c) shown in Ref.\cite{nakajima.wang.15}. 
The lattice parameter $a$ first decreases up to $p\sim8$~GPa and then starts to increase. 
The lattice constant $c$ decreases in a whole range of pressures.
A sharp enlargement of parameter $a$ and sudden reduction of parameter $c$ point out that the T--cT phase transition occurs at $p=14$~GPa. 
The measured x-ray diffraction spectra taken at room temperature indicate an abrupt change of lattice parameters around $16$~GPa. 
Such discrepancy between the measured and calculated pressure of the isostructural transition is not surprising taking into account that the calculated lattice constants correspond to temperature 0~K.  
Above the phase transition, parameter $a$ decreases again with pressure.

The unusual modifications of lattice parameters with pressure are induced by the changes of interatomic distances [Fig.~\ref{fig.struc}(b)]. 
Especially, the relation between the As--As interlayer distance and the distances of Fe--As and Fe--Fe in the FeAs plane seems to be crucial.  
Results of detailed analysis of the crystal geometry (Fig.~\ref{fig.struc}) are presented below:

\begin{enumerate}[(i)]
\item The shortest interatomic distance is that between the Fe and As atoms within one FeAs layer [Fig.\ref{fig.struc}(b)]. Its length, 2.32~\AA, is slightly smaller than the sum of Fe and As covalent radii indicating strong Fe--As bonds within the layers. 
With increasing pressure the Fe--As distance decreases very slowly.
\item The second shortest distance between neighboring Fe atoms, which initially equals 2.68~\AA, decreases with increasing pressure achieving minimum around 10~GPa.
Then it grows up around the phase transition and under higher pressure it starts slow decreasing again.
\item The interlayer distance between two As atoms [labeled as As-As in Fig.\ref{fig.struc}(b) and shown as a blue dashed line in Fig.\ref{fig.struc}(a)] equals to 4.25~\AA~ at zero pressure and drops dramatically to 2.94~\AA~ under compression to 15~GPa.
In cT phase, the increasing pressure leads to further shortening of this distance, however, the rate of distance decrease is notably slower.
\item The interplanar As--Fe--As bond angle [marked in Fig.\ref{fig.struc}(b) as $\alpha$] shows a nonmonotonous dependence on pressure and changes abruptly around the phase transition [Fig.~\ref{fig.struc}(c)]. Further increasing of pressure leads to only small changes of this angle.
\item The internal parameter $h_{\text{As}}$, denoting a distance between As--atoms and Fe--atom planes (a half of the FeAs layer width), first increases and then decreases with increasing pressure, showing the abrupt reduction at the phase transition [Fig.~\ref{fig.struc}(b)].
Additional increase of the pressure leads to only small decrease of its value.
\item Changes of the lattice constants [Fig.~\ref{fig.struc}(d)] with increasing pressure leads to simultaneous change of the volume of the tetragonal crystallographic cell and volume of the FeAs layer [Fig.~\ref{fig.struc}(e)]. 
At the transition point both values change discontinuously as a consequence of the collapse.
\item
Above the isostructural phase transition, all interatomic distances slowly decrease with an approximately linear dependence on pressure.
\end{enumerate}

The sharp decrease of the As--As distance with pressure is 
the common feature of compounds belonging to 122 family and exhibiting the T--cT phase transition.  
The distance between interlayer arsenic atoms decreases with pressure reaching a value that is lower than a sum of the van der Waals radii in the cT phase.  
For this reason the tetragonal collapse observed in the 122 compounds under pressure is discussed taking into account the direct bond formation  between interlayer As atoms or the overlap of arsenic $4p_z$ orbitals ~\cite{yildrin.09,kasinathan.schmitt.11,stavrou.chen.15}. 
At the same time, the strong in-plane Fe--As and Fe--Fe bonds become weaker and consequently the system changes its characteristic two-dimensional character to the three-dimensional collapsed phase at high pressure.

\begin{figure}[!t]
\centering
\includegraphics[width=\linewidth]{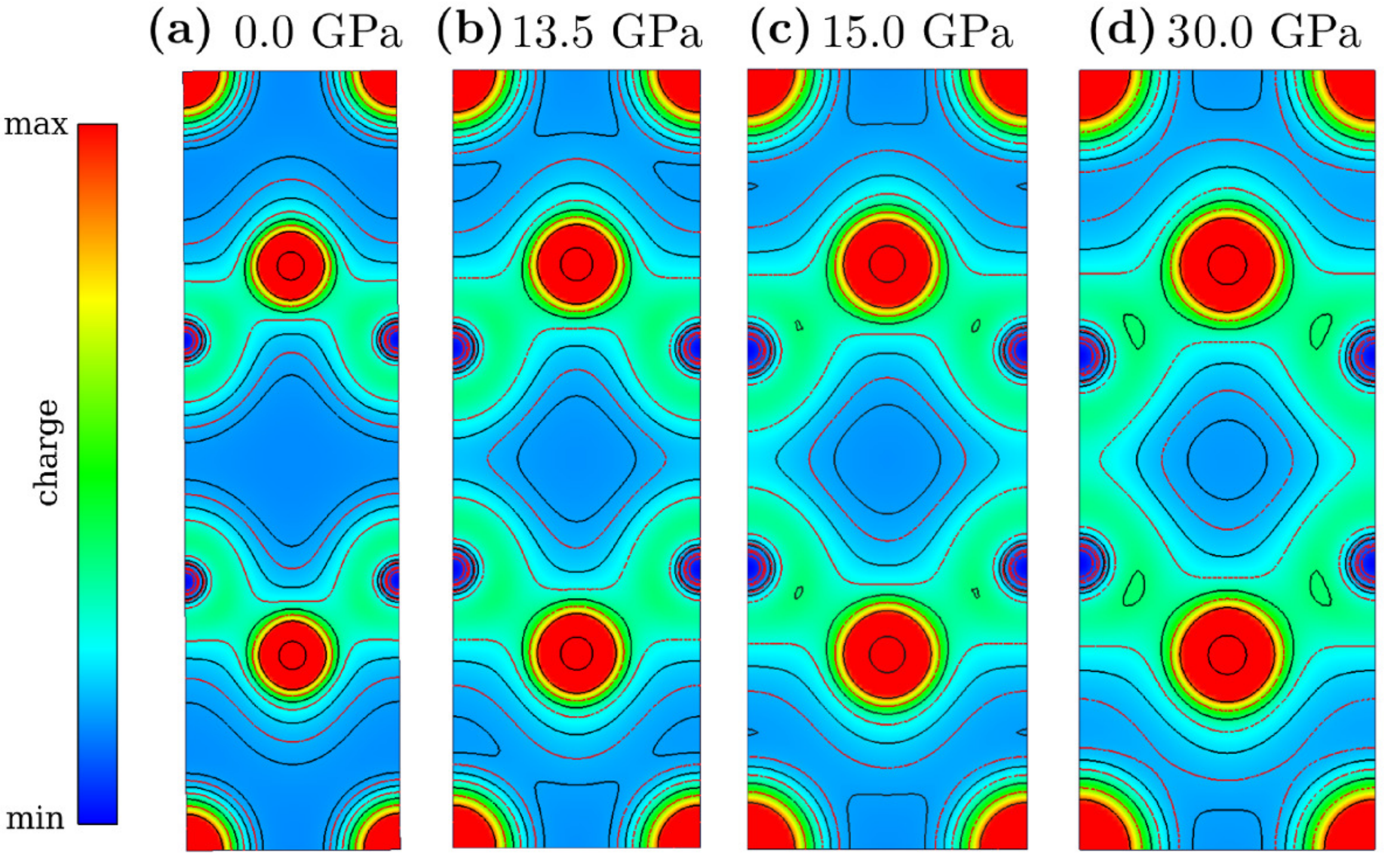}
\caption{
\label{fig.isocharge}
Contour maps of the spatial distribution of charge density in KFe$_{2}$As$_{2}$ in plane passing through interlayer As atoms calculated for four hydrostatic pressures. 
The densities from $\text{min} = 0$ to $\text{max} = 0.345$~e/\AA$^{3}$ are presented using contours plotted in logarithmic mode. 
}
\end{figure}

\begin{figure}[!b]
\centering
\includegraphics[width=\linewidth]{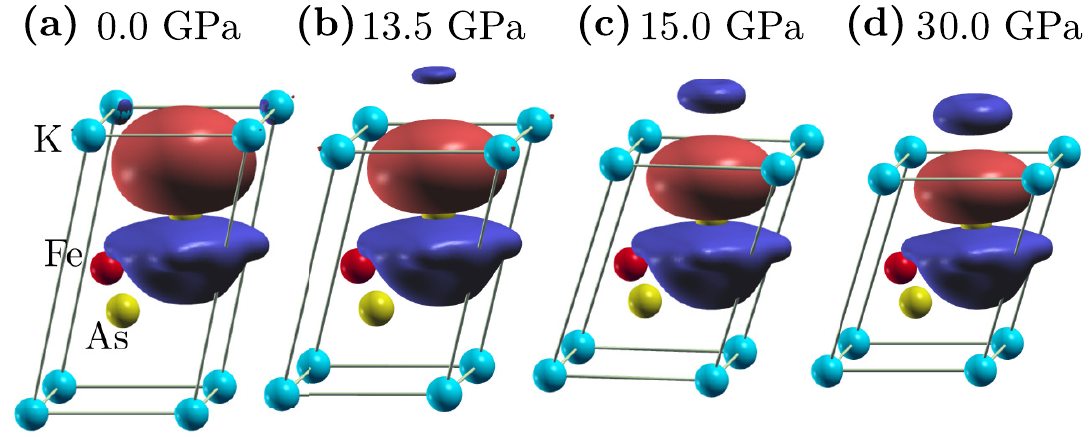}
\caption{
\label{fig.pzwannier}
$p_{z}$-like maximally localized Wannier orbital localized around As atoms.
For the simplicity only orbital originating from one As atom is shown. 
The second $p_z$-like orbital can by obtained by the mirror point symmetry.
Red and blue colors of the orbital correspond to different signs of the wave function. 
The primitive unit cell of K122 is shown for a comparison.
}
\end{figure}

The evolution of charge density distribution with pressure in KFe$_2$As$_2$ presented in $a$--$c$ plane passing through the As atoms (Fig.~\ref{fig.isocharge}) confirms previous findings. 
At zero pressure [Fig.~\ref{fig.isocharge}(a)] the strong covalent As--Fe bonds are characterized by the high charge density between these atoms while the weakly bonded neighboring FeAs planes are separated by the regions with a low charge density.  
The isocharge lines plotted in logarithmic mode (as implemented in {\sc VESTA} software) demonstrate the significant differences of densities.
Due to the very low charge density between two As atoms the large As--As distance is easily shortened by the imposed pressure until the T--cT transition pressure is achieved and the charge density is substantially higher.
Finally, at 30 GPa the charge density between interlayer As atoms is just as high as that between atoms belonging to the single FeAs layer.

In the orbital representation, the formation of As--As bonds is related to the overlap between the $p_{z}$ orbitals of As atoms from neighboring FeAs layers.
In Fig.~\ref{fig.pzwannier} we show the evolution of $p_z$-like MLW orbital centered at one As atom.
With increasing pressure the shape of this orbital is changing.
We can observe the emergence of an additional part of the orbital located outside the primitive cell.
Since the 122 structure exhibits the mirror point symmetry, the $p_z$ orbital associated with the nearest As atom (in the same unit cell) exists with the opposite phase. 
Initially focused in one unit cell [Fig.~\ref{fig.pzwannier}(a)], the orbitals begin to play a more important role in the bonding between the neighboring primitive cells along the $z$ direction.
As we can see, in contrary to the behavior of the isocharge surface at $p=0$~GPa, the $p_{z}$-like Wannier orbitals show a precursor of bonding between two neighboring cells along the $z$ direction below the structural phase transition.

\begin{figure}[!b]
\centering
\includegraphics[width=\linewidth]{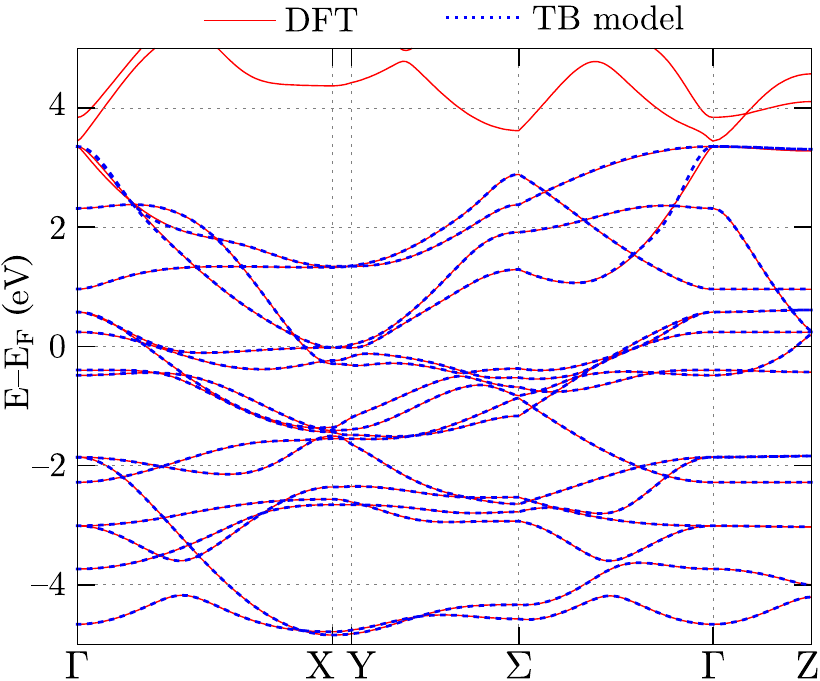}
\caption{
\label{fig.band0}
A comparison of the band structures of K122 obtained from DFT calculations (red solid lines) and tight binding model in maximally localized Wannier orbitals (blue dotted lines). 
Results in the absence of the hydrostatic pressure.
}
\end{figure}

\subsection{Electronic structure}
\label{sec.electrons}

As we mentioned above, for a reproduction of the electron band structure of the K122 compound, we fitted the TBM using the MLW orbitals basis.
To construct $16$ bands in the TBM, we take the Fe 3$d$ orbitals and As 4$p$ orbitals~\cite{cao.hirschfeld.08}.
A comparison of the band structure obtained within the {\it ab initio} (DFT) calculations and that extracted from the TBM (in the absence of the external pressure) is shown in Fig.~\ref{fig.band0}.
Around the Fermi level and far below, the model very well reproduces the band structure obtained within the DFT.
As we can see, the used model is more adequate than the previous ones, e.g. the 5-orbital/10-band model~\cite{suzuki.usui.11,li.li.12}.
For each pressure, we fit the parameters of the TBM, independently.

\begin{figure}[!t]
\centering
\includegraphics[width=\linewidth]{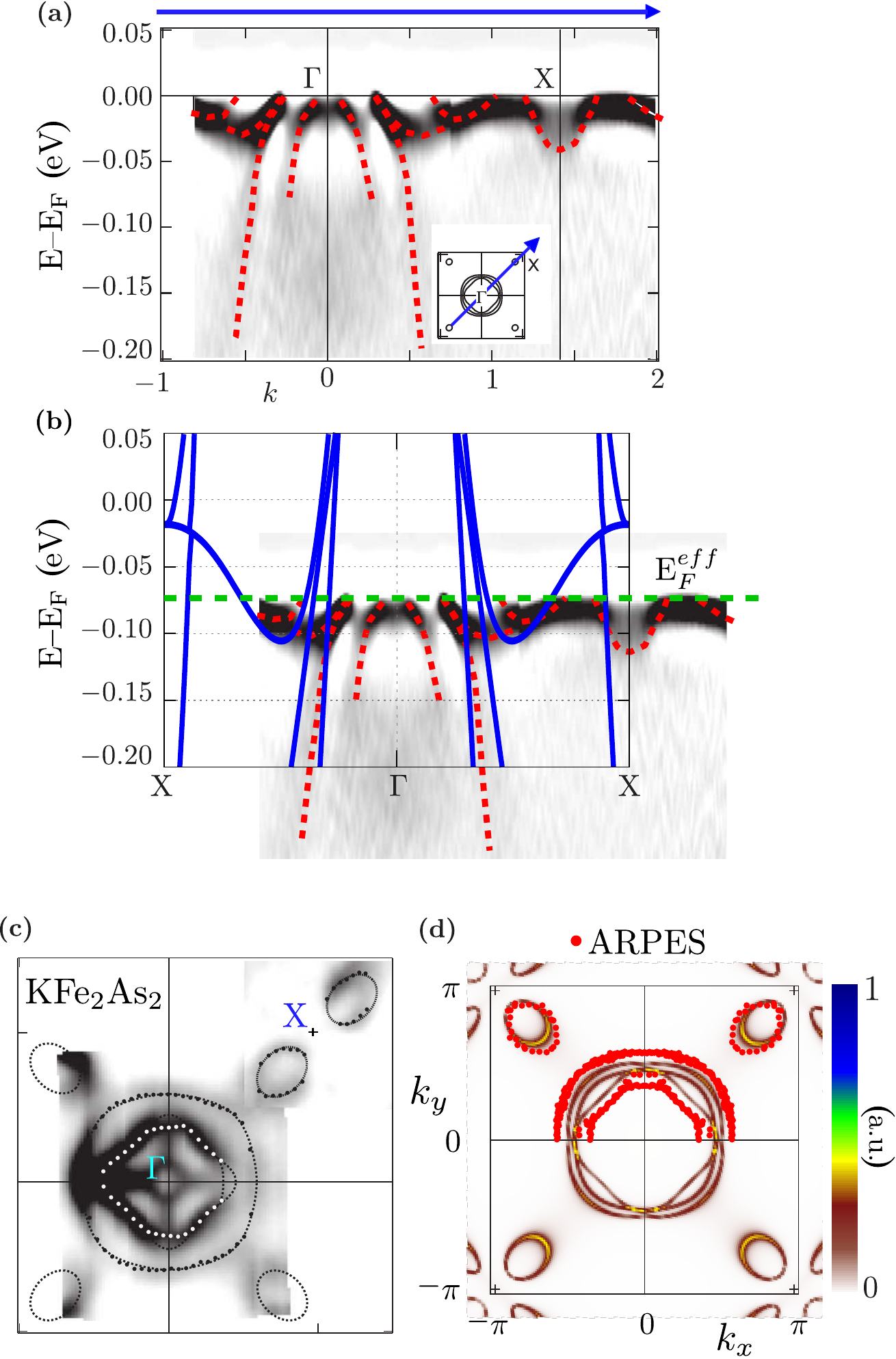}
\caption{
\label{fig.arpes}
Comparison of experimental ARPES results and DFT calculations.
(a) ARPES band structure along the $\Gamma$--X direction around the Fermi level.
(b) Comparison of the DFT results (blue line) with ARPES spectra from panel (a) (background and red line).
The DFT band structure is shifted with respect to experimental results by approximately 75~meV to higher energy (cf. blue and red lines). The green line denotes the Fermi level obtained from the ARPES.
(c) The Fermi surface at $\Gamma$--X plane obtained by ARPES.
(d) Comparison of the calculated spectral function $\mathcal{A} ( {\bm k} , \omega )$ at the effective (shifted) Fermi level $\omega = E^{eff}_{F}$ (background) with ARPES data (red dots).
ARPES data are adopted from Ref.~\cite{yoshida.nishi.11}.
}
\end{figure}

The obtained TBM defines the hopping integrals between different orbitals in several neighboring unit cells in every direction (precisely to all cells included in the supercell built of $13 \times 13 \times 13$ unit cells, while the initial cell is located at the center of the supercell).
It should be noted that near the structural phase transition, hopping integral changes its value due to larger overlap between orbitals along the $z$ axis. 
The largest modifications are found for the $p_{z}$ orbitals localized on two neighboring As atoms along the $z$ direction [shown as blue dashed line in Fig.~\ref{fig.struc}(a)].
For this bond the hopping integral changes its value from $1.394$~eV to $1.901$~eV.
This change is associated with the emergence of stronger As--As interlayer bonding~\cite{yildrin.09,kasinathan.schmitt.11,stavrou.chen.15}, described in the previous section.

The spectra measured by the angle resolved photoemission spectroscopy (ARPES) showed the existence of three hole pockets centred at the $\Gamma$ point and four small hole pockets around the X point~\cite{sato.nakayama.09,yoshida.nishi.11,tarashima.kurita.13,yoshida.ideta.14}.
A comparison of these results with the DFT calculations is presented in Fig.~\ref{fig.arpes}.
To achieve a consistency of these results with our DFT calculations it is necessary to shift theoretical results of approximately 75 meV to lower energies [cf. blue and red lines Fig.~\ref{fig.arpes}(b)].
A similar comparison can be done for the FS obtained by ARPES [Fig.~\ref{fig.arpes}(c)].
To do this, we have calculated the spectral function $\mathcal{A} ( {\bm k} , \omega )$ at the ``new'' shifted Fermi level $\omega = E_{F}^{eff}$.
The obtained result is shown as a color map in Fig.~\ref{fig.arpes}(d).
As we can see, the structure of shifted bands found by the DFT calculations very well reproduces the shape and size of the experimental FS.
Such a small shift of the Fermi level can be treated effectively as doping that changes the average number of particles by $\delta n \sim 0.075$~\cite{wang.kreisel.13}.
Here, it should be noted that such small doping does not change qualitatively the obtained electronic structure. 
This fixed ``shift'' of the band energies was also applied in the remaining
calculations carried out for the crystal under pressure.
We should also note that real samples may include some small deviations from the ideal stoichiometry, which are very difficult to observe experimentally.

\begin{figure}[!b]
\centering
\includegraphics[width=\linewidth]{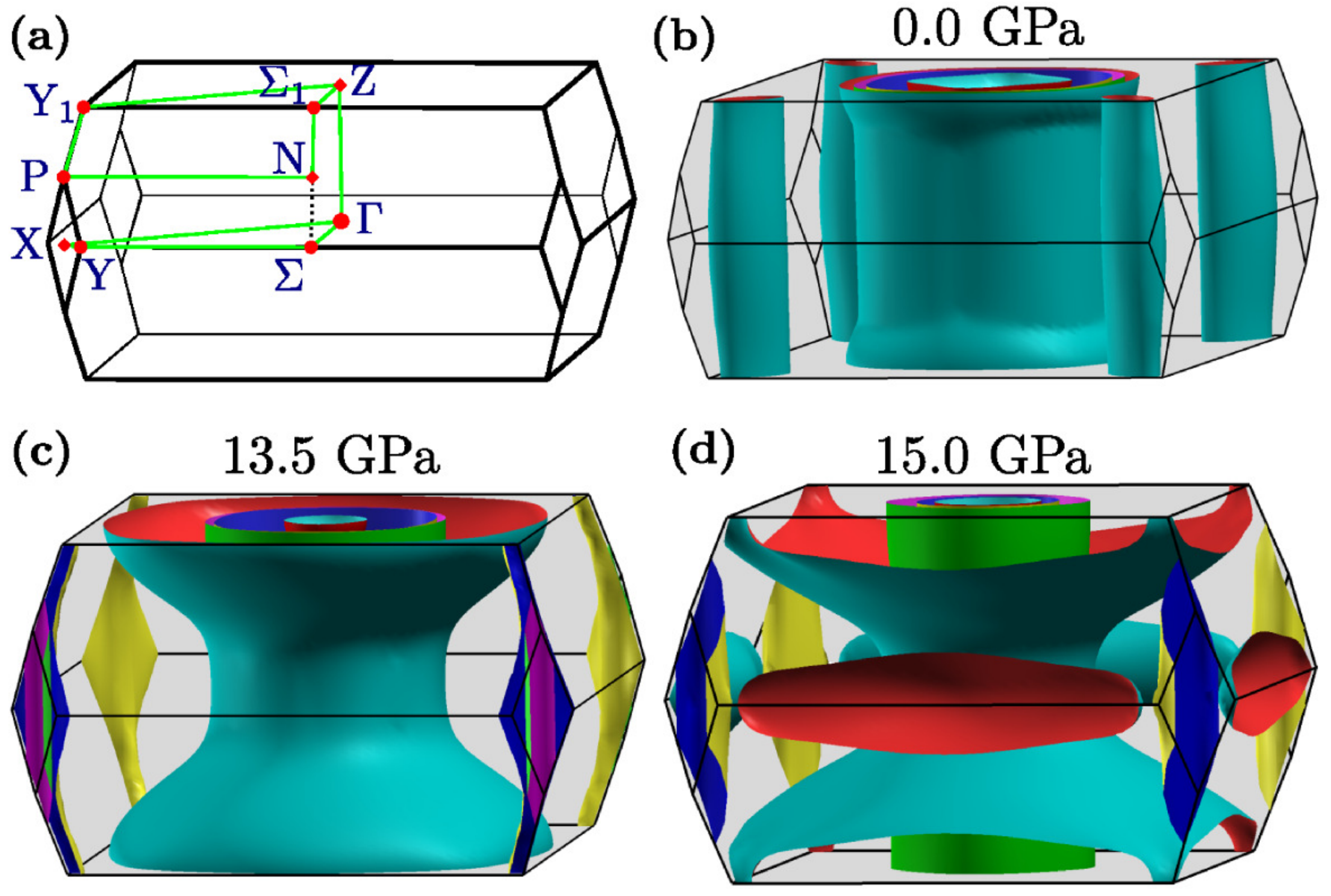}
\caption{
\label{fig.fs}
(a) The first Brillouin zone of the tetragonal structure ({\it I4/mmm}) and its high-symmetry points.
The Fermi surface of K122 at ambient pressure (b), before the structural transition (c), and after the structural phase transition (d).
The image was rendered using {\sc xCrysDen} software~\cite{xcrysden}.
}
\end{figure}

\begin{figure*}[!t]
\centering
\includegraphics[width=0.8\linewidth]{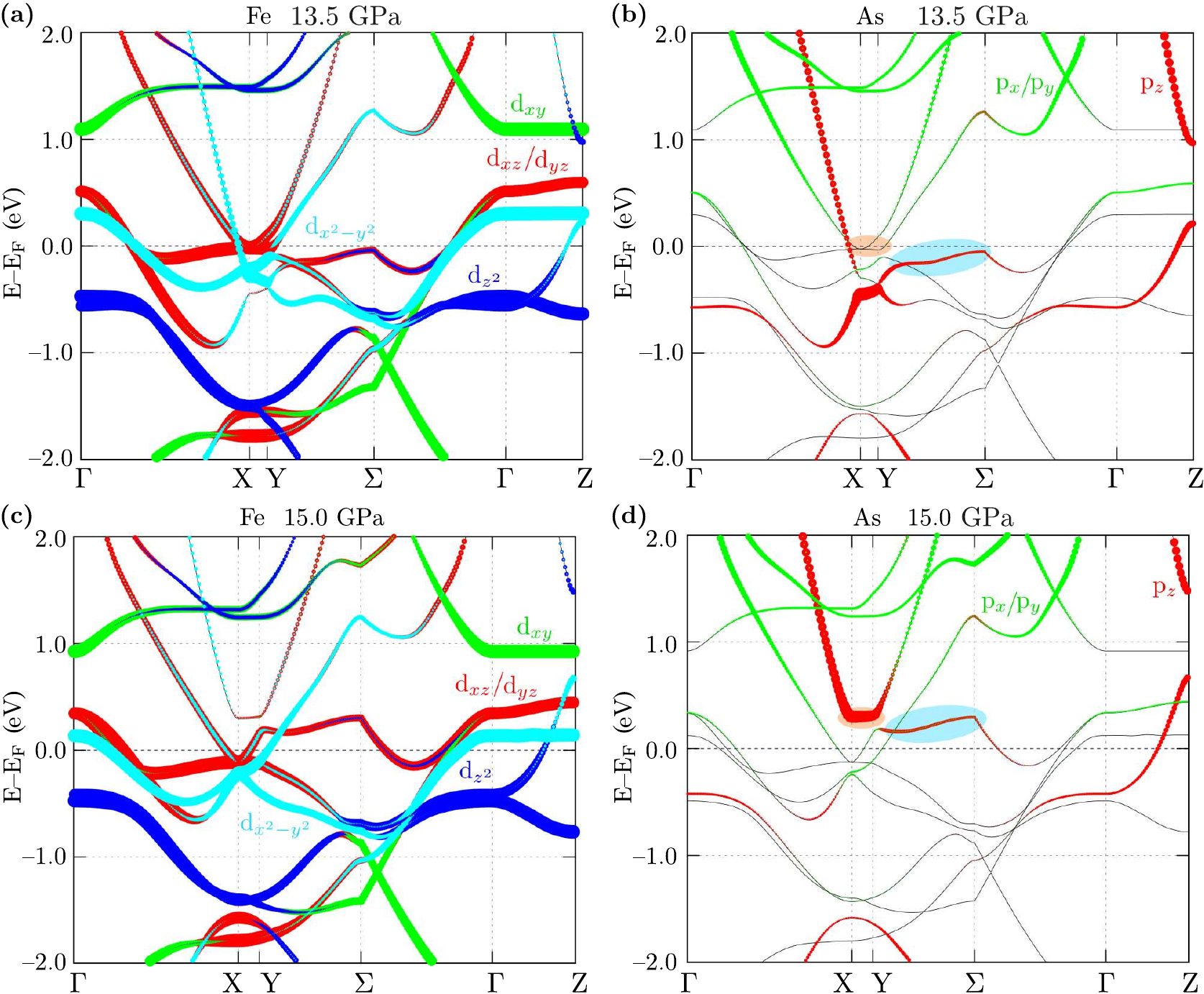}
\caption{
\label{fig.orbit}
Electron band projections on Fe $d$ orbitals (left column) and As $p$ orbitals (right column).
Sizes of dots correspond to the value of the contribution of a given orbital, while their colors correspond to their type (as labeled).
The results obtained below (top row, $p=13.5$~GPa) and above (bottom row, $p=15.0$~GPa) the structural phase transition (in the T and cT phases, respectively).
}
\end{figure*}

One  should also have in mind that the DFT calculations for IBSC, e.g., Refs.~\cite{mazin.johannes.08,singh.du.08,kuroki.onari.08,zabidi.abukassim.09,terashima.kimata.10,miyake.nakamura.10,thomale.platt.11,suzuki.usui.11,skornyakov.anisimov.14}, often give the band structure shifted with respect to the experimental ARPES data~\cite{evtushinsky.inosov.09,kordyuk.12,richard.sato.11,borisenko.evtushinksy.15,yoshida.ideta.14}.
Moreover, from a comparison of the band structure obtained from the ARPES and DFT calculations, we can observe not only the shift of the Fermi level, but also different {\it effective masses} $m^{\ast}$, i.e., different $1 / m^{\ast} \propto [ \partial^{2} E_{\bm k} / \partial {\bm k}^{2} ]_{{\bm k} = {\bm k}_{F}}$.
The experimental value of the effective electron mass $m^{\ast}/m_{e}$ is reported in the range from $6$ to $20$~\cite{terashima.kimata.10,mizukami.kawamoto.16,eilers.gube.16}.
This discrepancy is a consequence of the incorrect description of correlations in IBSC by the DFT approach.
This is a well-know problem, which has been discussed many times in the literature~\cite{subedi.zhang.08,qazilbash.hamlin.09,aichhorn.biermann.10} and can be solved by the band renormalization~\cite{murai.suzuki.18} or the combination of DFT with the dynamic mean field theory (DMFT)~\cite{haule.shim.08,li.ji.12,ferber.foyevtsova.12,backes.guterding.14,backes.jeschke.15,derondeau.bisti.17,yang.yin.17} .
A better inclusion of the correlation effects gives more realistic band structures, ordering of magnetic moments, effective masses as well as Fermi surfaces~\cite{yin.haule.11}.
However, in the present work, the main objective is the influence of pressure on properties of K122, therefore, the purpose of the Fermi level shift is to obtain the proper band structure at zero pressure.

The FS of the IBSC have a characteristic cylindrical shape~\cite{borisenko.zabolotnyy.10,kordyuk.12,kordyuk.zabolornyy.13,derondeau.bisti.17}. 
The FS calculated for K122 at $p=0$~GPa using the shifted Fermi level,  shown in Fig.~\ref{fig.fs}(b), agrees with the experimental results obtained by ARPES~\cite{sato.nakayama.09,yoshida.ideta.14}.
It is formed by three hole-like bands, which participate in a creation of three pockets in the middle of the Brillouin zone (BZ) along the $\Gamma$--Z line and four small pockets located close to the corners of the BZ (at the X--P line).

With increasing pressure, a shape of the FS is modified both in the neighborhood of the BZ boundary at $k_{z} = \pi / c$ and close to the X--P--Y$_{1}$ boundary [cf. Fig.~\ref{fig.fs}(b) and Fig.~\ref{fig.fs}(c)]. 
The pocket observed close to the X point at $p=0$~GPa is shifted to the zone boundary and changes its character from hole-like to the electron-like at higher pressures.
This type of changes in the FS of K122 has been already discussed~\cite{nakajima.wang.15} and reported also for other 122 family members~\cite{gonnelli.daghero.16,tresca.profeta.17}.

A further increase of pressure causes the occurrence of a new structure  in the $\Gamma$--X ($k_{z} = 0$) plane of the FS when the system passes through the structural transition pressure [Fig.~\ref{fig.fs}(d)]. 
This phenomenon is known as Lifshitz transition induced by the pressure~\cite{lifshitz.60}.
To discuss it more precisely we calculate the orbital-projection band structure.  
Details of these calculations are presented in Appendix~\ref{sec.app.orbital_proj}.
The results of numerical calculations are shown in Fig.~\ref{fig.orbit} for the Fe $d$ orbitals and As $p$ orbitals (left and right columns, respectively).
K ions do not have considerable contribution in the presented energy range and do not play any role in the construction of the TBMs. 
In the case of the Fe $d$ orbitals, the structural transition modifies the bands close to the Fermi level mainly along the X--Y--$\Sigma$ line [cf. Fig.~\ref{fig.orbit}(a) and (c)]. 
Some parts of the $d_{xz}/d_{yz}$ bands located just below the Fermi level in the T phase are shifted above it in the cT phase.
Similar behavior is observed for the As $p$ orbitals, where the modifications of the bands in this part of the BZ are even more pronounced [cf. Fig.~\ref{fig.orbit}(b) and (d)]. 
Two bands associated with the $p_{z}$-like orbitals (marked by the red and blue ovals) located below the Fermi level are shifted by increasing pressures above it.
The band projections on the $p_{x}$- and $p_{y}$-like orbitals do not change.

The Lifshitz transition associated with the T--cT phase transition in K122 is mainly caused by shift of bands lying along X--Y--$\Sigma$ line.  
These bands initially located below Fermi level are moved above it. 
Another change observed above the transition pressure is the enhanced band energy dispersions along the $\Gamma$--Z. 
It is interesting to note that filling of all bands is changed at the Lifshitz transition.
This is experimentally observed, e.g., in Hall coefficient $R_{H}$, which is proportional to the inverse of the total filling $1/n$~\cite{madsen.singh.06}.
For K122 the Hall coefficient changes its sign at the structural transition~\cite{tafti.juneaufecteau.13,ying.tang.15}.

The Lifshitz transition leads to the modification of the FS topology. 
It also changes contributions of given orbitals, which forms the FS.
A perfect nesting between electron and hole pockets of the FS~\cite{chubukov.efremov.08} reported in many families of the IBSC~\cite{winiarski.samselczekala.12,bao.qiu.09,winiarski.samselczekala.13,rodriguez.stock.11,ciechan.winiarski.13,dai.15,mou.kong.16} occurs in the absence of the external pressure.
However, in K122 at pressures below the isostructural transition the perfect nesting is not observed, 
and one cannot expect an existence of any magnetic order or nematic phases~\cite{chubukov.khodas.16,li.su.17}.
The situation looks different for the cT phase. 
In this case, the Lifshitz transtion (i.e., the emergence of new pockets of the FS around the boundary of the BZ) leads to the FS shape~\cite{nakajima.wang.15,tresca.profeta.17} resembling that observed in the magnetic YFe$_{2}$Ge$_{2}$, what has been suggested also in Ref.~\cite{chen.semeniuk.16}. 
Similar properties can be also expected in the case of YRu$_{2}$Ge$_{2}$, which present the
same Fermi surface structure~\cite{chajewski.samselczekala.18}.
The other indication of the presence 
of the magnetic order in the cT phase is the vanishing superconducting state.  Its disappearance can be a consequence of the competition between magnetism and superconductivity.

\begin{figure}[!b]
\centering
\includegraphics[width=\linewidth]{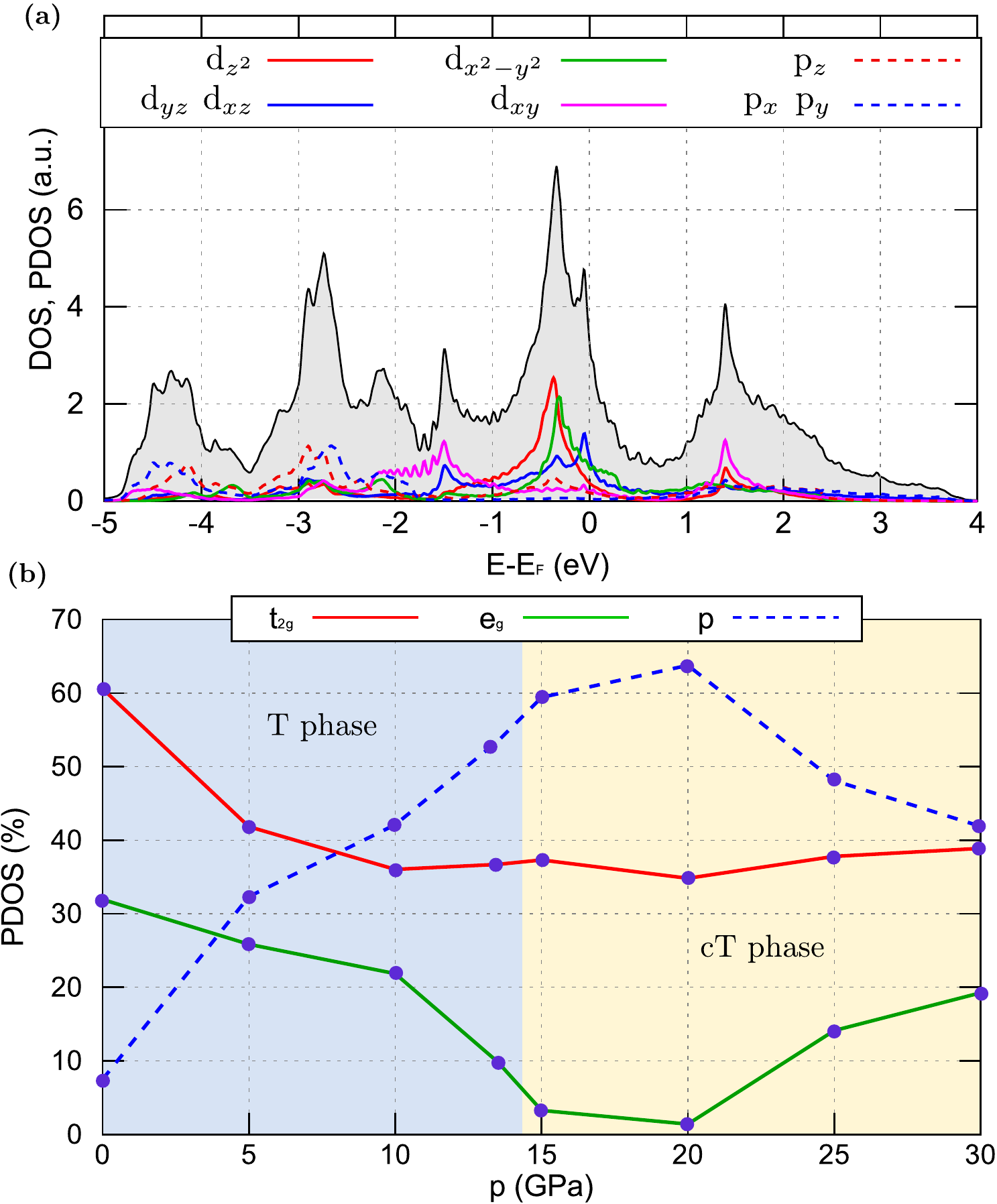}
\caption{
\label{fig.el_dos}
(a) Total electronic density of states (black solid line and the shadow) and partial electronic DOS projected onto orbitals of Fe and As atoms (color labeled lines as indicated) at $p=0$ GPa.
(b) Contributions of given types of the orbitals to the total DOS at the Fermi level as a function of the external pressure. 
$t_{2g}$ denotes the contribution of $d_{xz}$, $d_{yz}$ and $d_{xy}$ orbitals of the Fe atoms together, $e_g$ denotes the contribution of $d_{z^2}$ and $d_{x^2-y^2}$ orbitals of Fe atoms together, whereas $p$ is the joint contribution of all $p$-like orbitals of As atoms. 
}
\end{figure}

\subsection{Superconducting properties}
\label{sec.supercond}

IBSC are characterized by the unconventional superconductivity~\cite{scalapino.12,chubukov.12,hosono.kuroki.15}, i.e., the anisotropic gap (symmetries other than {\it s-wave}, typically {\it s$_{\pm}$} symmetry~\cite{mazin.singh.08,kuroki.onari.08,ikeda.arita.10,wang.berlijn.15}) and mediated by antiferromagnetic spin fluctuations~\cite{mazin.singh.08,wang.lee.11,hirschfeld.korshunov.11,suzuki.usui.11,li.li.12,bang.stewart.17}.
In the relation to this, in the absence of the pressure, the experimental results for K122 show that this material is characterized  by the {\it d-wave}-type gap with nodal lines~\cite{thomale.platt.11,okazaki.ota.12}, what was reported, e.g. in magnetic penetration-depth ~\cite{hasimoto.serafin.10,watanabe.yamashita.14}, thermal conductivity ~\cite{watanabe.yamashita.14,reid.tanatar.12,dong.zhou.10,fukazawa.yamada.09} or nuclear quadrupole resonance~\cite{fukazawa.yamada.09} measurements.
However, with increasing pressure, the gap symmetry can be changed from {\it d} to {\it s$_{\pm}$}~\cite{guterding.backes.15}.

In the low pressure regime, K122 exhibits a sudden change of behavior of critical temperature $T_{c}$, from its initial decrease with pressure to an increase above pressure $p_{c}= 1.75$~GPa~\cite{tafti.juneaufecteau.13}.
Moreover, the universal V--shape temperature-pressure phase diagrams of K122, Rb122 and Cs122 have been reported~\cite{tafti.ouellet.15}.
In addition, the critical magnetic field initially decreases with the external pressure, whereas above $p_c$ it increases with pressure~\cite{tarashima.kihou.14}. 
Experimental analyses of the FS in the presence of weak pressure do not provide any additional information about superconducting properties~\cite{grinenko.schottenhamel.14}.
At higher pressures ($p>4.5$ GPa), the critical temperature decreases and superconductivity disappears above 11 GPa~\cite{wang.matsubayashi.16}.

\begin{figure}[!t]
\centering
\includegraphics[width=\linewidth]{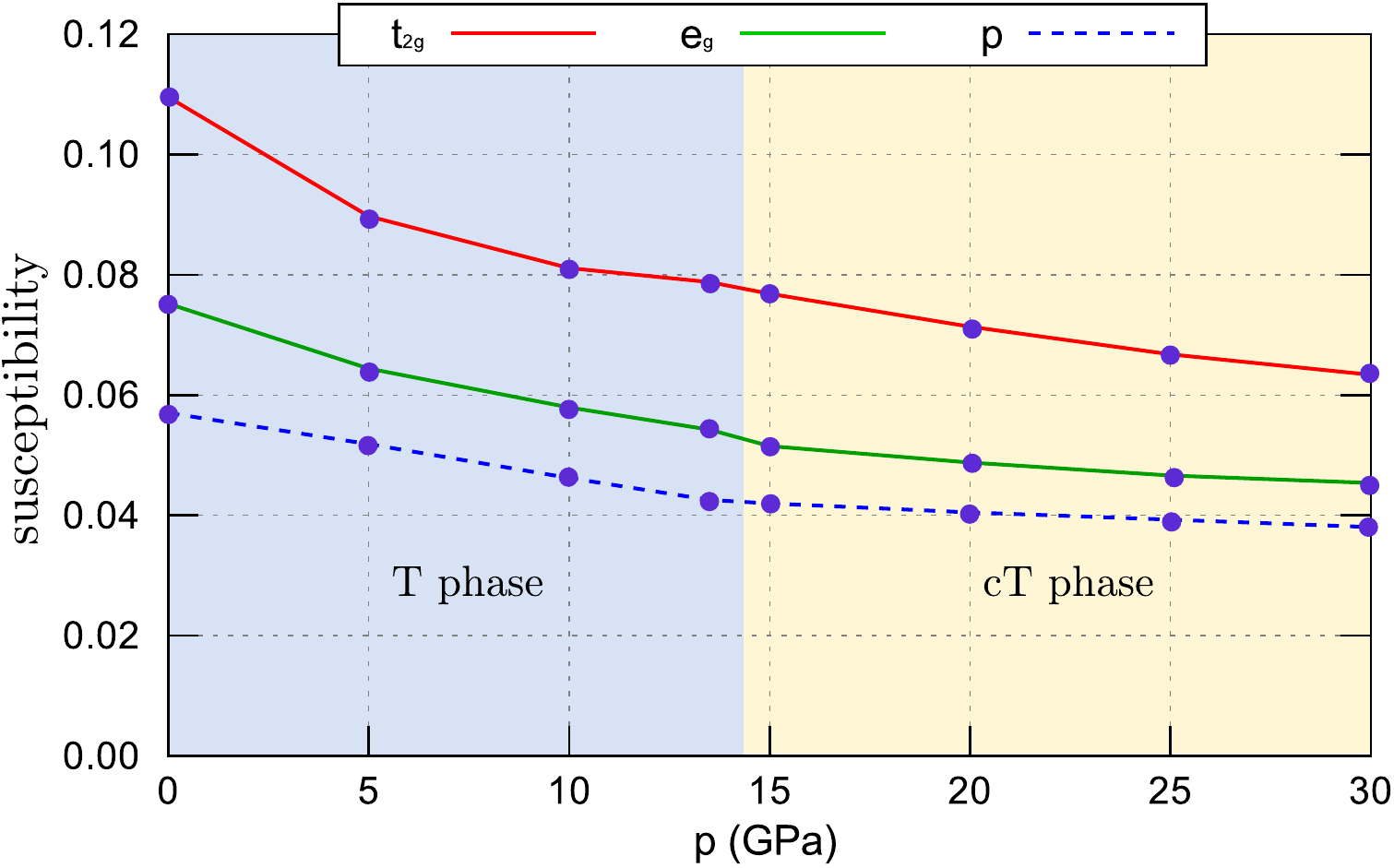}
\caption{
\label{fig.cps}
The pressure dependence of the superconducting intraband susceptibility originating from  particular groups of the orbitals, [as labeled, denotations as in Fig.~\ref{fig.el_dos}(b)].
}
\end{figure}

Superconducting properties of 122 systems result from the filling of $d$-orbitals ~\cite{mizukami.kawamoto.16}.
By analogy to Ba122, one can expect that strong electron pairing should be present in the $3d_{xz/yz}$ Fe orbitals~\cite{evtushinsky.zabolotnyy.14}.
In this context, the orbital contribution to the total DOS at the FS [presented in Fig.~\ref{fig.el_dos}(b)] can be an important indicator of the role of particular orbitals in the superconducting state.
In the absence of external pressure, the local maximum in the DOS is located near the Fermi level [Fig.~\ref{fig.el_dos}(a)].
The results of our calculations presented in Fig.~\ref{fig.el_dos}(b) show the contributions of $d$ and $p$ orbitals to the total DOS.
As we can see, with increasing pressure an influence of the $t_{2g}$ orbitals at the Fermi level decreases initially from $60\%$ to $37\%$. 
Next, at larger pressures, its value is approximately constant, while the contribution of the $e_{g}$ ($p$) orbitals decreases (increases) dramatically around the phase transition.

Similar analysis can be performed by using the Cooper pairs susceptibility (for details of the method see in e.g. Ref.~\cite{ptok.kapcia.17}). 
This quantity tells us about a tendency of the system to realize superconductivity.
As we can see from Fig.~\ref{fig.cps}, similarly as discussed previously, a role of $d$ orbitals decreases with increasing pressure.
If we assume that only $d_{xz/yz}$ orbitals on Fe atoms are responsible for superconductivity (cf. Ref.~\cite{evtushinsky.zabolotnyy.14}), the decreasing value of susceptibility in these orbitals is consistent with the diminution of superconductivity under pressure observed experimentally.

\subsection{Dynamical properties}
\label{sec.phonons}

In this section, we analyze the influence of pressure and structural phase transition on lattice dynamics in K122.
Details of the calculation method are described in Appendix~\ref{sec.app.phonons}.
\begin{figure} [!t]
\centering
\includegraphics[width=\linewidth]{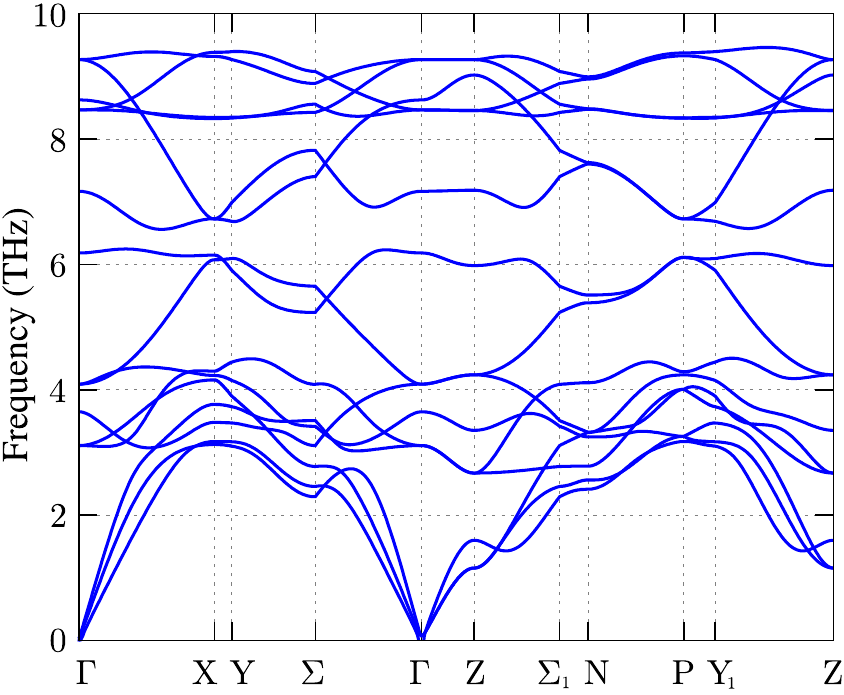}
\caption{\label{fig.ph_dc}
Phonon dispersion curves of K122 along the high-symmetry directions of the first Brillouin zone at $p=0$~GPa.
}
\end{figure}
\begin{figure} [!b]
\centering
\includegraphics[width=\linewidth]{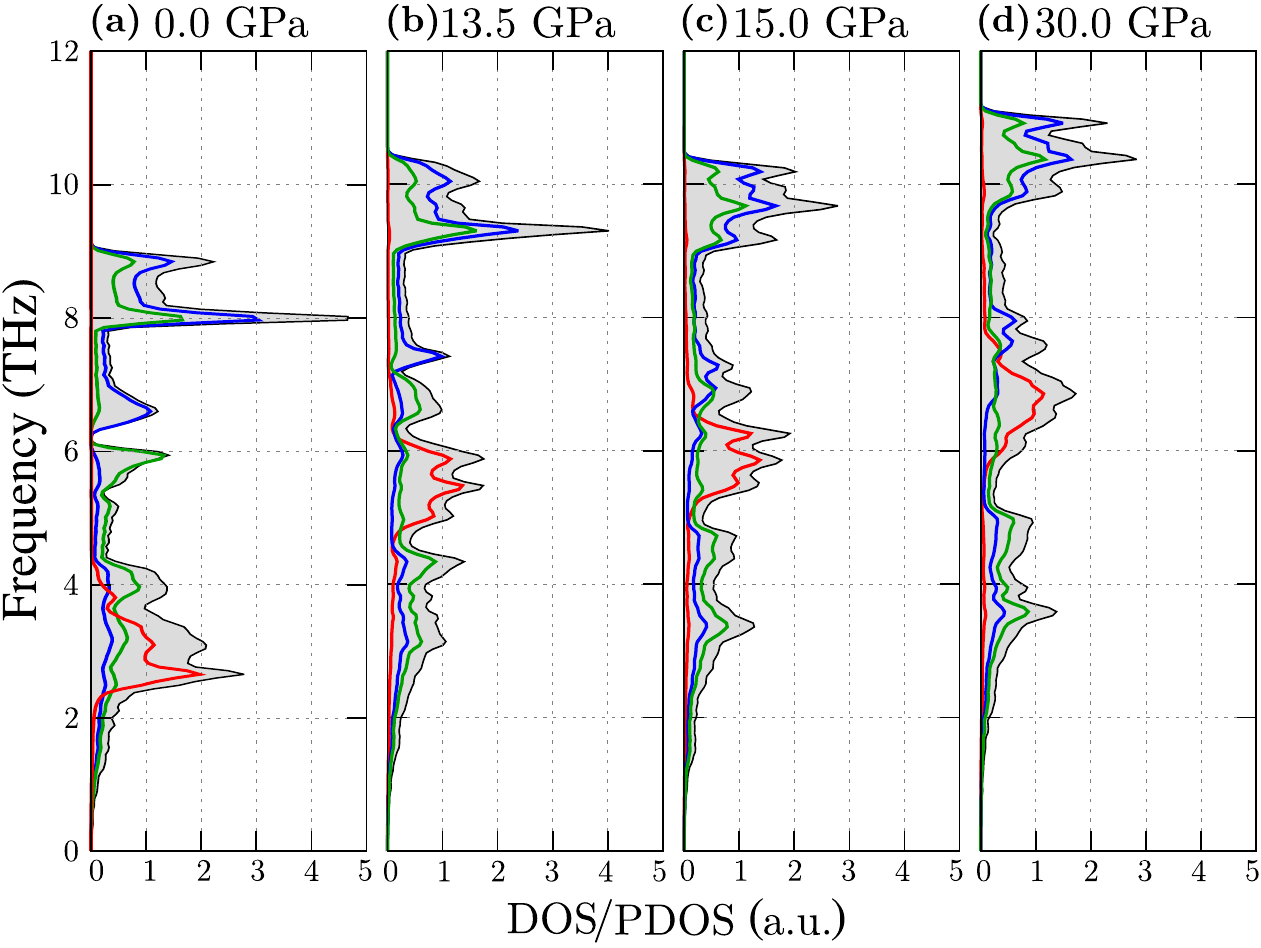}
\caption{\label{fig.ph_dos}
Total phonon DOS (shaded grey area) and partial phonon DOS (red, blue, green lines correspond to the K, Fe and As atoms, respectively) for several values of the hydrostatic pressure.}
\end{figure}
\begin{figure*}[!t]
\centering
\includegraphics[width=\linewidth]{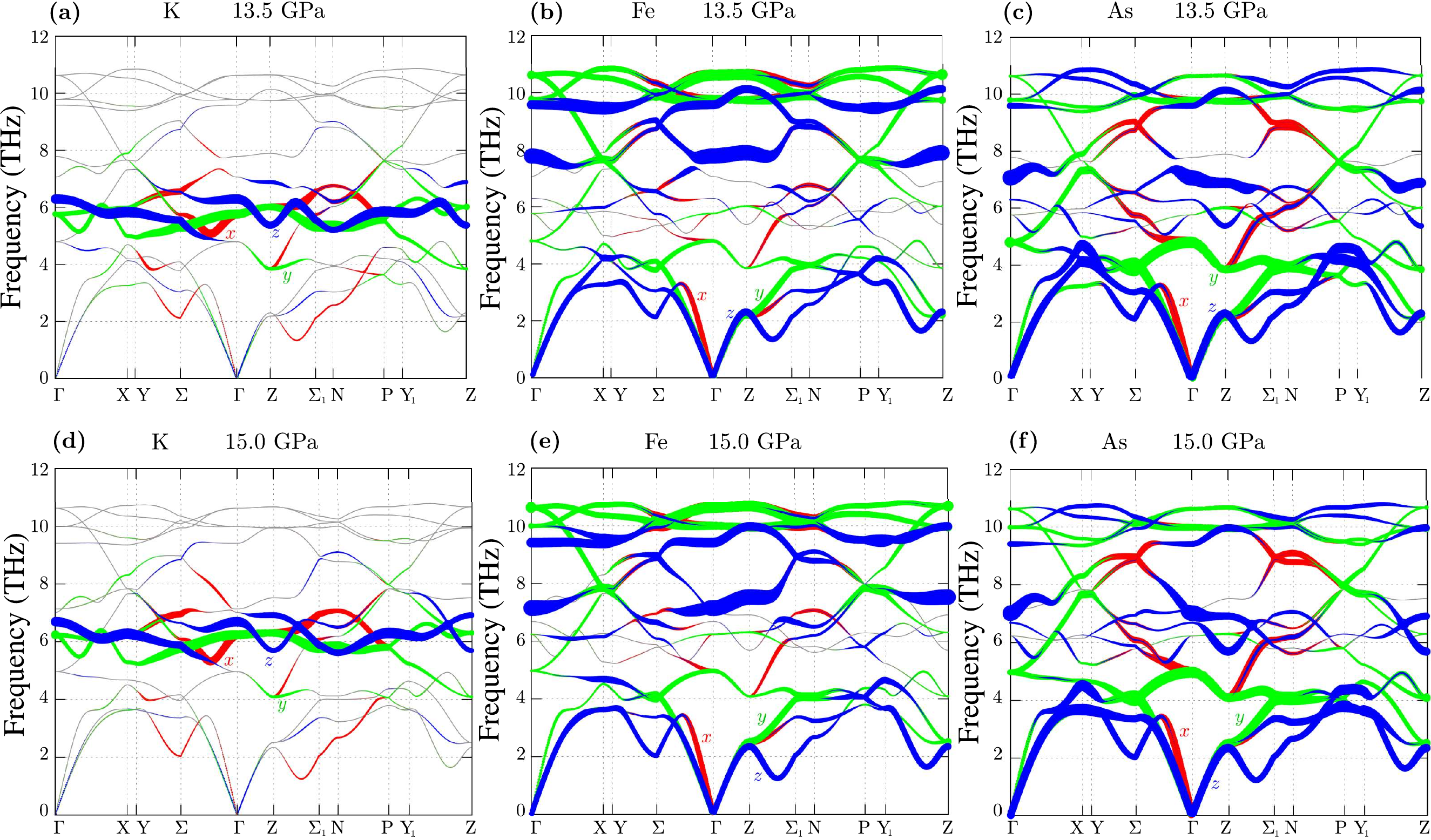}
\caption{\label{fig.phon_press}
The phonon band structure projections on atoms (columns as labeled) and directions (line colors) below (top row, $p=13.5$~GPa) and above (bottom row, $p=15.0$~GPa) the structural phase transition.
Sizes of dots correspond to the value of the contribution of oscillations of a given atom, while their colors determines the direction of that oscillations.
Red, green, and blue colors correspond to oscillations in $x$, $y$, and $z$ directions, respectively.
}
\end{figure*}
The phonon dispersion relations calculated at $p=0$~GPa are presented in Fig.~\ref{fig.ph_dc}.
At any point in the Brillouin zone, there are $15$ vibrational normal modes as it is expected for a crystal with $5$ atoms in the primitive unit cell. 
Phonon frequencies cover the range up to $9$ THz including a small gap around $6$~THz. 
The total phonon density of states and individual contributions of atoms to the total DOS are presented in Fig.~\ref{fig.ph_dos}. 
At zero pressure the vibrations of potassium atoms dominate at low frequencies between $2$ and $4$~THz (red lines), 
while the vibrational frequencies of Fe and As atoms (blue and green lines, respectively) cover a whole frequency range.
With increasing pressure all phonon frequencies shift to higher energies with the largest changes observed for the vibrations of K atoms. A small gap in the phonon DOS observed at $p=0$, disappears at higher pressures. 
Additionally, the intensities of peaks creating the highest-frequency part of spectra dramatically change between 13.5 and 15 GPa.
Overall, the spectra obtained below [Fig.~\ref{fig.ph_dos}(a) and (b)] and above [Fig.~\ref{fig.ph_dos}(c) and (d)] the structural phase transition are clearly distinct due to strong changes in the interatomic distances and interactions. The lack of any imaginary (soft) modes in the phonon DOS indicates a dynamical stability of K122 up to $p=30$ GPa.

Next, we follow through the phonon dispersion relations at $13.5$~GPa and $15.0$~GPa (Fig.~\ref{fig.phon_press}) to analyze in detail the changes in lattice dynamics induced by the phase transition.
Colors of lines correspond to the phonon projection on a given direction ($x$, $y$, and $z$ directions correspond to red, green, and blue color, respectively), while the line thickness corresponds to the projection value.  
In this way, we present the contribution of vibrations of a chosen atom to a selected phonon branch in a given direction. 
The dispersion curves taken at both pressures are quite similar, however, some differences are noticeable. 
The K atom vibrations contribute strongly to optical branches in a narrow range of frequencies around $6$~THz and their frequencies increase slightly due to the phase transition. 
The Fe atoms contribute mainly to the highest-energy optical modes as well as to the acoustic modes.
The vibrations of Fe atoms along the $z$ direction are particularly strong at well defined branches started from the $\Gamma$ point at $7.78$~THz.  
In spite of the contraction of lattice constant $c$ the Fe--Fe interatomic distance increases so the frequency of this mode decreases to $7.14$~THz.
This effect is also very strong for the Fe modes propagating along the $\Gamma$--Z direction. 
The strongest As vibrations are observed at the $\Gamma$, X, Y, and P high-symmetry points.
The frequencies of these modes are slightly lower than frequencies of the most intensive Fe phonons.
The largest changes induced by the phase transition are observed for the As modes around $7$~THz with polarization along the $z$ direction.

\begin{figure}[!t]
\centering
\includegraphics[width=\linewidth]{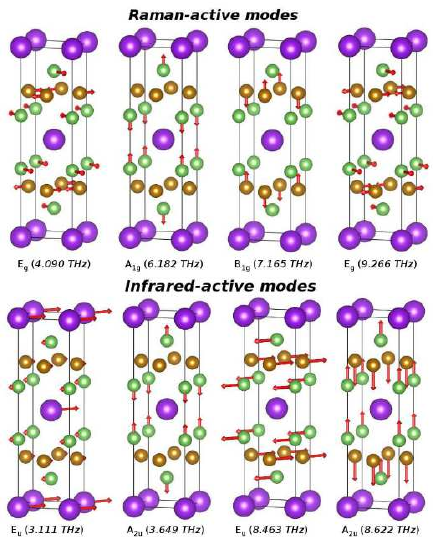}
\caption{
Schematic illustration of Raman and infrared active modes in KFe$_2$As$_2$. 
In the parenthesis the frequency of each mode is given. 
\label{fig.ph_mody}
}
\end{figure}

Frequencies at the $\Gamma$ point can be verified experimentally using the Raman and infrared (IR) spectroscopy. 
Group theory analysis indicates that the $12$ optical modes form the following irreducible representations at the Brillouin zone center, $\Gamma_{op} = 2E_u^{(2)} + 2A_{2u} + 2E_{g}^{(2)} + A_{1g} + B_{1g}$
where $A_{1g}$, $E_g$, and $B_{1g}$ modes are Raman active and $A_{2u}$, $E_u$ are infrared active. 
In agreement with the dimension of these representations, all $E$ modes are doubly degenerate at the $\Gamma$ point. 
Atomic displacements corresponding to the Raman and IR active modes in K122, schematically presented in Fig.~\ref{fig.ph_mody}, are consistent with those observed previously in KCo$_2$Se$_2$~\cite{wdowik.jaglo.15}.
There are no Raman modes associated with the vibrations of the K-sublattice. 
In contrast, vibrations of K atoms dominate in two of the IR active modes $E_u$ and $A_{2u}$. The first is related to the strong atomic movements of K atoms along the $z$ direction and the other one with vibrations in perpendicular plane whereas the amplitudes of As and Fe atoms are about one tenth of K displacements.

\begin{figure}[!b]
\centering
\includegraphics[width=\linewidth]{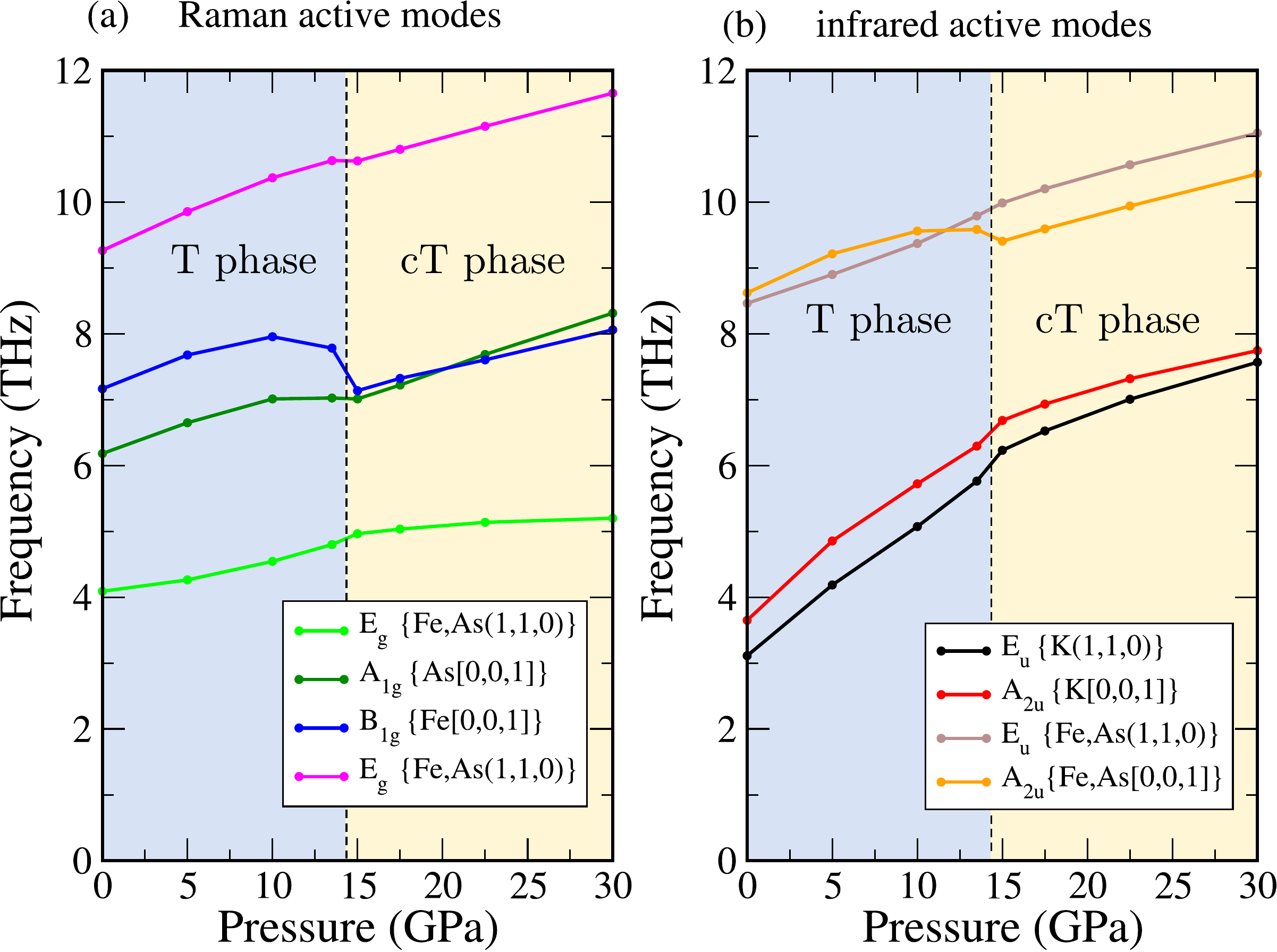}
\caption{
Evolution of Raman active (a) and infrared active modes (b) frequencies with pressure. 
\label{fig.omega}
}
\end{figure}

In Fig.~\ref{fig.omega}, the pressure dependence of each of eight optic modes at the $\Gamma$ point is presented. 
To characterize these modes, the symmetry, names of atoms that are mainly involved in, and the direction of atomic movements are given. 
For example, $B_{1g}~\{Fe[0,0,1]\}$ identifies vibrations of Fe atoms along the $z$ direction described by the $B_{1g}$ symmetry and $A_{1g}~\{Fe, As(1,1,0)\}$ represents vibrations dominated by movements of As and Fe in the $xy$ plane.

At zero pressure, the two lowest modes are the IR phonons related with the strong movements of K atoms in the $(1,1,0)$ plane and along the $[0,0,1]$ direction.
Their frequencies systematically increase with pressure, however, the slopes of lines become steeper in close vicinity of the structural phase transition, i.e. between $13.5$~GPa and $15$~GPa [Fig.~\ref{fig.omega}(b)].
Similarly, pronounced changes in frequencies are observed for all modes in this region of pressure.
The frequency of the $A_u$ mode, which involves vibrations of Fe and As atoms along the $z$ direction shows a nonmonotonic behavior.
First, it increases up to 10~GPa, then drops between 10-15~GPa, and increases again above the phase transition.
A similar behavior is found for two Raman modes $A_{1g}$ and $B_{1g}$ representing vibrations of As and Fe atoms, respectively, along the $z$ direction.
As it is shown in Fig.~\ref{fig.phon_press}, these modes correspond to the most intensive phonons propagating along the $\Gamma$--X and $\Gamma$--Z directions. 
They can be promising objects for experimental and theoretical investigations of phonons or even phonon-electron interactions as it was shown in the case of Ba$_{1-x}$K$_x$Fe$_2$As$_2$ ($x=0.28$, superconducting T$_{c}$=29~K), Sr$_{1-x}$K$_x$Fe$_2$As$_2$ ($x=0.15$, T$_{c}$=29~K), and nonsuperconducting Ba122 single crystals~\cite{rahlenbeck.sun.09}, Ba$_{0.6}$K$_{0.4}$Fe$_2$As$_2$~\cite{bohm.kemper.12} or Fe$_{1+y}$Te$_{1-x}$Se$_x$~\cite{um.subedi.12}.

There are no experimental data on phonon dispersion relations or phonon DOS for the undoped K122 compound. 
Neutron scattering measurements of the phonon DOS in Sr$_{0.6}$K$_{0.4}$Fe$_2$As$_2$ (T$_{c}$=32~K) and Ca$_{0.6}$Na$_{0.4}$Fe$_2$As$_2$ (T$_{c}$=28~K) compared with the Ba122 compound show that the doping or exchange of alkali atoms affect mainly the low and intermediate frequency range of the spectra~\cite{mittal.su.08}. The spectrum presented in Fig.~\ref{fig.ph_dos}(a) confirms this findings as the potassium atoms vibrate with low frequencies and in the high-frequency part of the spectrum only Fe and As atoms are involved. 
Experimental studies of the pressure dependence of vibrational frequencies in both the tetragonal and collapsed tetragonal phases have been performed for Ca122 using the inelastic x-ray and neutron scattering~\cite{mittal.heid.09} and 
more recently for Sr122 using the nuclear resonant inelastic x-ray scattering~\cite{wang.lu.16}.
It has been found that the phonon frequency shifts induced by pressure are well explained by changes in relevant bond lengths throughout the entire pressure range. Our calculations also indicate that nonmonotonic changes in the distance between two neighboring Fe atoms [Fig.~\ref{fig.struc}(b)] significantly influence the vibrational frequency of the Fe--Fe stretching $B_{1g}$ mode along the $z$ axis. 
As a result, the frequency softening is observed just a few GPa below the isostructural transition. 
On the other hand, phonons polarized in the (1,1,0) plane stiffen under the imposed pressure.
In general, in accordance with the previous studies \cite{mittal.heid.09,wang.lu.16}, the structural phase transition influences much stronger  phonon modes
polarized along the $z$ direction than those in the $xy$ plane.

Let us stress that the IBSC are unconventional superconductors in the sense that phonons likely do not play a dominant role in their superconductivity.
The calculations based on the density-functional perturbation theory lead to the conclusion that the total electron-phonon coupling is too weak to explain their high T$_{c}$~\cite{boeri.dolgov.08,mazin.singh.08,naidyuk.kvitnitskaja.14,boeri.calandra.10}, but strong enough to have a non-negligible effect on superconductivity, for instance, by frustrating the coupling with spin fluctuations and inducing the order parameter nodes~\cite{boeri.calandra.10}. Therefore, the analysis performed in this section may be important for full understanding of the superconducting properties of K122.

\section{Summary}
\label{sec.summary}

Using the {\it ab initio} calculations, we have studied the influence of pressure and the isostructural phase transition on the properties of the KFe$_{2}$As$_{2}$ superconductor. 
The obtained lattice parameters as functions of pressure show a very good agreement with the experimental data.

To explore the electronic and superconducting properties, we have built the  $16$-orbital tight biding model within the maximally localized Wannier orbitals. 
The compatibility between  the resulted electronic structure and that obtained from DFT calculations has been mainly taken into account in the selection of TBM parameters.

We have discussed the mechanism leading to the phase transition from the tetragonal phase to the collapsed tetragonal phase,
which is associated with an enhancement of bonding between As atoms along the $z$ direction.
We have revealed the precursor of the phase transition in a form of an additional part of the Wannier orbital on the As atoms.

We have investigated changes of the Fermi surface induced by pressure and analyzed the Lifshitz transition that occurs when the system passes through the isostructural phase transition.  
The detailed studies of the pressure dependence of As-p and Fe-d bands
have demonstrated that the new structure observed in the $\Gamma$--X plane of the FS results from
the shift of some bands located below the Fermi level to energies above it.

To specify the role of the particular orbitals in the superconducting state, the orbital contributions to the total DOS have been also considered. 
We have found that in the T phase the contributions of $e_{g}$ and $t_{2g}$ orbitals to the total DOS at the Fermi energy decrease with pressure achieving at the transition pressure, the values significantly smaller then rapidly growing the $p$ orbital contribution.   
We have also investigated the effect of pressure on the pairing susceptibility.
Modification of the electronic structure and the calculated pairing susceptibility allow us to conclude that the imposed pressure should reduce $T_{c}$ in agreement with the experiment.

Finally, we have analyzed the changes in lattice dynamics induced by pressure and
found a significant modification of phonon spectra, especially the modes corresponding to vibrations of Fe and As atoms along the $z$ directions.
They show the anomalous, non-monotonic dependence of phonon frequency on pressure close to the isostructural phase transition. 
This result can be verified experimentally using the Raman and infrared spectroscopy.

\begin{acknowledgments}
The authors are thankful to Jan \L{}a\.{z}ewski, Pawe\l{} Jochym, Andrzej M. Ole\'{s}, and Krzysztof Parlinski for the very fruitful discussions and comments.
This work was supported by the National Science Centre (NCN, Poland) under grants UMO-2016/21/D/ST3/03385 (A.P.), UMO-2017/25/B/ST3/02586 (M.S. and P.P.), and UMO-2017/24/C/ST3/00276 (K.J.K.).
\end{acknowledgments}

\appendix

\section{Tight binding model}
\label{sec.app.tbmodel}

The TBM in the MLW functions (orbitals)~\cite{marzari.mostofi.12} can be express in the form:
\begin{eqnarray}
\mathcal{H} = \sum_{{\bm R}{\bm R}',\mu\nu,\sigma} t_{{\bm R},{\bm R}'}^{\mu\nu} c_{{\bm R}\mu\sigma}^{\dagger} c_{{\bm R'}\nu\sigma} ,
\end{eqnarray}
where $t_{{\bm R},{\bm R}'}^{\mu\nu}$ is the hopping elements between orbitals $\mu$ and $\nu$ localized at sites indicate by vectors ${\bm R}$ and ${\bm R}'$. 
Here, $c_{{\bm R}\mu\sigma}^{\dagger}$ ($c_{{\bm R}\mu\sigma}$) is the creation (annihilation) operator [in the orbital (Wannier) basis] of an electron with spin $\sigma$ at orbital $\mu$ of ${\bm R}$ atoms.
From the {\it ab initio} (DFT) calculation, the hopping integrals can be found by the {\sc Wannier90} program~\cite{mostofi.yates.08,mostofi.yates.14}.
The Hamiltonian in the momentum space takes the form:
\begin{eqnarray}
\label{eq.ham_momW} \mathcal{H} = \sum_{{\bm k}\mu\nu\sigma} \mathbb{H}_{\mu\nu} ( {\bm k} ) c_{{\bm k}\mu\sigma}^{\dagger} c_{{\bm k}\nu\sigma} ,
\end{eqnarray}
where 
\begin{eqnarray}
\label{eq.ham_mom} \mathbb{H}_{\mu\nu} ( {\bm k} ) = \sum_{\delta} \exp \left( i {\bm k} \cdot \delta \right) t_{\delta}^{\mu\nu} .
\end{eqnarray}
Here, we take $\delta = {\bm R} - {\bm R}'$ -- the distance between two atoms in the real space and $t_{\delta}^{\mu\nu} \equiv t_{{\bm R},{\bm R}'}^{\mu\nu}$.
The band structure can be found by a diagonalization of the matrix~(\ref{eq.ham_momB}).
Then, the Hamiltonian takes the diagonal form:
\begin{eqnarray}
\label{eq.ham_momB} \mathcal{H} ( {\bm k} ) = \sum_{n} E_{{\bm k}n\sigma} d_{{\bm k}n\sigma}^{\dagger} d_{{\bm k}n\sigma} ,
\end{eqnarray}
where $E_{F}$ denotes the Fermi level and $d_{{\bm k}n\sigma}^{\dagger}$ ($d_{{\bm k}n\sigma}$) is the new creation (annihilation) operator [in the band (Bloch) basis] of the electron with momentum ${\bm k}$ and spin $\sigma$.
Additionally, the band number is given by $n$. 
Relations between the operators in the Wannier basis and the Bloch basis are given by the unitary transformation:
\begin{eqnarray}
\label{eq.unitary} d_{{\bm k}n\sigma} = \sum_{\nu} \mathcal{U}_{{\bm k}\nu n} c_{{\bm k}\nu\sigma} .
\end{eqnarray}
Matrix $\mathcal{U}$ is composed by the eigenvectors of the matrix $\mathbb{H}_{\mu\nu} ( {\bm k} )$~(\ref{eq.ham_mom}) and diagonalizes (transforms) the Hamiltonian from the form~(\ref{eq.ham_momW}) into~(\ref{eq.ham_momB}).

\section{Orbital projections of electronic band structure}
\label{sec.app.orbital_proj}

The density of states (DOS) from its definition is given as:
\begin{eqnarray}
\label{eq.dos} 
\rho ( \omega ) & = & - \frac{1}{\pi} \sum_{{\bm k}\nu\sigma} \langle\langle c_{{\bm k}\nu\sigma} | c_{{\bm k}\nu\sigma}^{\dagger} \rangle\rangle \\
\nonumber 
 &= & - \frac{1}{\pi} \sum_{{\bm k}n\sigma} \langle\langle d_{{\bm k}n\sigma} | d_{{\bm k}n\sigma}^{\dagger} \rangle\rangle = \sum_{{\bm k}} \mathcal{A}({\bm k},\omega),
\end{eqnarray}
where $\mathcal{A}({\bm k},\omega)$ is the spectral function.
Substituting~(\ref{eq.ham_momB}) and~(\ref{eq.unitary}) to~(\ref{eq.dos}) we can find an exact expression of the DOS:
\begin{eqnarray}
\nonumber \rho ( \omega ) = \sum_{\nu} \left\lbrace \sum_{{\bm k}n\sigma} | \mathcal{U}_{{\bm k}\nu n} |^{2} \delta \left( \omega - E_{{\bm k}n\sigma} \right) \right\rbrace = \sum_{\nu} \rho_{\nu} ( \omega ) . \\
\end{eqnarray}
Coefficients $| \mathcal{U}_{{\bm k}\nu n} |^{2}$ denote projections of the $n$-th electronic (spin $\sigma$) band (at momentum ${\bm k}$) with energy $E_{{\bm k}n\sigma}$ onto the Wannier orbital $\nu$. 
Here, $\rho_{\nu} ( \omega )$ plays the role of the partial electronic DOS, projected onto $\nu$ orbital.

\section{Dynamical matrix}
\label{sec.app.phonons}

The dynamical properties of the system are given by the dynamical matrix:
\begin{eqnarray} 
\nonumber D_{\alpha\beta}^{jj'} ( {\bm q} ) = \frac{ 1 }{ \sqrt{m_{j}m_{j'}} } \sum_{n} \Phi_{\alpha\beta} ( j0,j'n ) \exp \left( i {\bm q} \cdot {\bm R}_{n} \right) , \\
\label{eq.dyn_mat}
\end{eqnarray}
where ${\bm q}$ is the phonon wave vector and $m_{j}$ denotes the mass of the atom $j$.
Here, $\Phi_{\alpha\beta} ( j0,j'n )$ is the force constants tensor ($\alpha$ and $\beta$ is the direction index: $x$, $y$, and $z$) between atoms $j$ and $j'$ localized in the initial ($0$) and $n$-th primitive unit cell, which can be find from the {\it ab initio} method using, e.g., the Parlinski--Li--Kawazoe method~\cite{phonon1}.

Energy spectrum of vibrations with a wave vector ${\bm q}$ is given as an eigenproblem of dynamical matrix~(\ref{eq.dyn_mat}):
\begin{eqnarray}
\sum_{j'\beta} D_{\alpha\beta}^{jj'} ( {\bm q} ) e_{\varepsilon {\bm q} \beta j'} = \left[ \Omega_{\varepsilon{\bm q}} \right]^{2} e_{\varepsilon {\bm q} \alpha j} ,
\end{eqnarray}
where $\Omega_{\varepsilon{\bm q}}$ corresponds to the frequency of $\varepsilon$ branch phonon.
Here, $3N$-component eigenvector $e_{\varepsilon {\bm q} \alpha j}$ describes the polarization of the phonon with frequency $\Omega_{\varepsilon{\bm q}}$ and wavevector ${\bm q}$.
Additionally, from the definition, the chosen components of the polarization vector $e_{\varepsilon {\bm q} \alpha j}$ describe the oscillations of the atom $j$ in the direction $\alpha$.

Similarly as for electrons, we can define the total phonon DOS for frequency $\Omega$, which is given by:
\begin{eqnarray}
\rho ( \Omega ) = \sum_{\varepsilon{\bm q}} \delta \left( \Omega - \Omega_{\varepsilon{\bm q}} \right) .
\end{eqnarray}
It can be expressed by the partial phonon DOS [denoted as $\rho_{\alpha j} ( \Omega )$]:
\begin{eqnarray}
\nonumber \rho ( \Omega ) = \sum_{\alpha j} \left\lbrace \sum_{\varepsilon{\bm q}} | e_{\varepsilon {\bm q} \alpha j} |^{2} \delta \left( \Omega - \Omega_{\varepsilon{\bm q}} \right) \right\rbrace = \sum_{\alpha j} \rho_{\alpha j} ( \Omega ) , \\
\end{eqnarray}
along directions $\alpha$ and/or atoms $j$.
As a consequence, a coefficient $| e_{\varepsilon {\bm q} \alpha j} |^{2}$ denotes the influence of $\varepsilon$ phonon with a wavevector ${\bm q}$ on to the oscillations of $j$ atom in $\alpha$ direction.

\bibliography{biblio}

\end{document}